\documentclass[aps,superscriptaddress,showpacs,11p]{revtex4}
\usepackage{graphicx,epsf,amssymb,amsmath,latexsym}
\newcommand{\be}{\begin{eqnarray}}
\newcommand{\ee}{\end{eqnarray}}
\def\ben{\begin{equation}}
\def\een{\end{equation}}
\def\bena{\begin{eqnarray}}
\def\eena{\end{eqnarray}}

\newcommand{\beq}{\begin{equation}}
\newcommand{\eeq}{\end{equation}}
\newcommand{\bqa}{\begin{eqnarray}}
\newcommand{\eqa}{\end{eqnarray}}

\begin{document}





\begin{flushright}
DAMTP-2005-116
\end{flushright}

%
\title{Magnetized Tolman-Bondi Collapse}

\vspace{.3in}

\author{Cristiano Germani}
\email{C.Germani@damtp.cam.ac.uk} \affiliation{D.A.M.T.P., Centre
for Mathematical Sciences, University of Cambridge, Wilberforce
road, Cambridge CB3 0WA, England}

\author{Christos G. Tsagas}
\email{tsagas@astro.auth.gr} \affiliation{Section of Astrophysics,
Astronomy and Mechanics, Department of Physics, Aristotle
University of Thessaloniki, Thessaloniki 54124, Greece}

\date{\today}

\vskip.3in

\begin{abstract}
We investigate the gravitational implosion of magnetized matter by
studying the inhomogeneous collapse of a weakly magnetized
Tolman-Bondi spacetime. The role of the field is analyzed by
looking at the convergence of neighboring particle worldlines. In
particular, we identify the magnetically related stresses in the
Raychaudhuri equation and use the Tolman-Bondi metric to evaluate
their impact on the collapsing dust. We find that, despite the low
energy level of the field, the Lorentz force dominates the
advanced stages of the collapse, leading to a strongly anisotropic
contraction. In addition, of all the magnetic stresses, those that
resist the collapse are found to grow faster.
\end{abstract}


\maketitle

\section{Introduction}
Magnetic fields are common features of almost all astrophysical
environments and stellar magnetism is a long established and very
active branch of astrophysics~\cite{P}. Nevertheless, analytical
studies of magnetic fields in strong gravity environments are less
developed. Most of the available work addresses the possible
gravitational effects on the Maxwell field (e.g.~see~\cite{BS} and
references therein) and relatively few look into the implications
of magnetic fields for gravitational collapse itself. Perhaps the
most intriguing result so far has been obtained by Thorne in his
analysis of Melvin's cylindrical magnetic universe~\cite{M}.
There, by developing the concept of `cylindrical energy', the
author reached the conclusion that ``a strong magnetic field along
the axis of symmetry may halt the collapse of a finite cylinder
before the singularity is reached''~\cite{Th}. The possible
support of the field against the gravitational collapse of massive
bounded systems was also studied in~\cite{AP}. That analysis led
to solutions of Einstein-Maxwell equations with no singularities
or event horizons, where the gravitational attraction is balanced
solely by magnetic stresses. Other work, however, indicated that
locally naked singularities may develop in some
cases~\cite{barrow}. Studies of contracting charge dust has
suggested that the fluid may "rebounce'', thus preventing
black-hole formation~\cite{novi}. It has been argued, on the other
hand, that a collapsing spherically symmetric charged dust will
inevitably produce naked singularities due to
shell-crossing~\cite{ori}. The latter indicates the intersection
of matter flow-lines along certain spacelike hypersurfaces.
Although these singularities are considered weak~\cite{N}, since
the curvature invariants and the tidal forces remain finite, their
appearance could also signal that a nonzero Lorentz force and
charged spherical collapse are physically incompatible. This
tentative `conjecture' is supported by the results of the present
paper, which focuses on the role of the magnetic Lorentz force
during the gravitational collapse of charged matter.

We consider the gravitational contraction of an inhomogeneous
spatially flat Tolman-Bondi model, filled with a pressureless
highly conductive fluid, and allow for a perturbing weak magnetic
field. This describes to a collapse of a weakly magnetized charged
plasma. The weakness of the field is measured by its contribution
to the total energy of the system. Our aim is to investigate the
fate of the contracting fluid in the presence of a nonzero and
without any symmetry constraints  Lorentz force. To the best of
our knowledge this question has not been addressed analytically.
Nonzero Lorentz force means that, despite the absence of fluid
pressure, the particle worldlines are no longer timelike
geodesics. As a result, we expect the initial spherical symmetry
of the collapse to break. We quantify the consequences of the
magnetic presence by looking at the convergence of neighboring
particle worldlines. This is done by analyzing the magnetic
contribution to the Raychaudhuri equation. The latter describes
the volume evolution of a given fluid element and has played a
fundamental role in numerous studies of gravitational collapse and
also in the formulation of the major singularity theorems
(e.g.~see~\cite{HE}). A relatively weak magnetic field contributes
to the Raychaudhuri equation primarily via its Lorentz force. The
benefit of our approach is that it identifies the effects of
Lorentz force on the collapsing matter directly. In particular,
the magnetic input splits up into a pair of stresses one of which
always supports against the gravitational pull of the matter.
Based on the weakness of the field, we ignore its backreaction on
the Tolman-Bondi metric and use the latter to evaluate the
aforementioned magnetic stresses. We find that the magnetic
presence triggers as range of effects with an overall impact that
depends on the specifics of the field in a rather involved way.
These effects can severely distort the spherical symmetry and
completely dominate the advanced stages of the Tolman-Bondi
collapse despite the low levels of the magnetic energy input.
Interestingly, of all the magnetic stresses, we found that those
supporting against the collapse grow faster and we have also
identified physically plausible magnetic configurations where this
happens. Although one should be very cautious before extrapolating
a linear result into the nonlinear regime, our analysis seems to
agree with earlier work on magnetized
collapse~\cite{Th}-\cite{ori}. Put together, these studies suggest
that a nonzero Lorentz force may be physically incompatible with
spherically symmetric collapse and that there might exist
situations where the support of the field can outbalance the total
gravitational attraction, at least along certain directions.

\section{Worldlines of magnetized matter}
Assume a general spacetime filled with a highly conductive perfect
fluid and allow for a magnetic field. High conductivity means that
there is no electric field and that the magnetic field is `frozen
in' with the matter. This is the well known MHD approximation
(e.g.~see~\cite{P}). In what follows we will investigate the
implications of the pure magnetic component of the Lorentz force
on the collapse of such a model. We will do so by testing the
convergence of the particle worldlines using the covariant
approach to general relativity~\cite{E}.

Covariantly, the dynamics of gravitational collapse is monitored
through the Raychaudhuri equation, which describes the volume
evolution of a self-gravitating fluid element. Consider a
congruence of timelike worldlines tangent to the 4-velocity field
$u_a$ (with $u_au^a=-1$) that follows the motion of the fluid.
Raychaudhuri's formula determines the evolution of
$\Theta=\nabla_au^a$, the scalar measuring the average contraction
(or expansion) between a pair of neighboring particle
worldlines~\cite{E}. In a magnetized environment we have~\cite{TB}
\begin{equation}
\dot{\Theta}+ {\textstyle{1\over3}}\Theta^2=-
{\textstyle{1\over2}}\left(\rho+3p+B^2\right)-
2\left(\sigma^2-\omega^2\right)+ {\rm D}^a\dot{u}_a+
\dot{u}_a\dot{u}^a\,,  \label{Ray}
\end{equation}
where $\rho$ and $p$ are respectively the energy density and
pressure of the fluid, $B^2=B_aB^a$ measures the energy density
and the isotropic pressure of the magnetic field ($B_a$),
$\sigma^2$ and $\omega^2$ are the respective magnitudes of the
shear and the vorticity associated with $u_a$ and
$\dot{u}_a=u^b\nabla_bu_a$ is the 4-acceleration. When the
right-hand side of the above is negative definite, an initially
converging family of worldlines will focus
(i.e.~$\Theta\rightarrow-\infty$) within a finite amount of
time~\cite{HE}. Thus, positive definite terms in the right-hand
side of (\ref{Ray}) will resist against further gravitational
contraction.

The magnetic contribution to the Raychaudhuri equation comes form
the energy density of the field which adds to the gravitational
pull of the matter and also from the magnetic input to the
4-acceleration. The latter satisfies the momentum-density
conservation law, which for a magnetized, highly conductive
perfect fluid takes the form~\cite{TB}
\begin{equation}
\left(\rho+p+{\textstyle{2\over3}}B^2\right)\dot{u}_a=-{\rm D}_ap-
\epsilon_{abc}B^b{\rm curl}B^c- \Pi_{ab}\dot{u}^b\,. \label{Euler}
\end{equation}
In the above ${\rm D}_a=h_a{}^b\nabla_a$ is the covariant
derivative operator orthogonal to $u_a$, with $h_{ab}$
representing the projection tensor (i.e.~$h_{ab}u^b$), while
$\Pi_{ab}=-B_{\langle a}B_{b\rangle}$ describes the anisotropic
pressure of the field~\footnote{Angled brackets indicate the
symmetric, trace-free part of projected second-rank tensors
(e.g.~$B_{\langle a}B_{b\rangle}=B_aB_b-(B^2/3)h_{ab}$, where
$h_{ab}=g_{ab}+u_au_b$ with $h_{ab}u^b=0$).}. The second term in
the right-hand side of Eq.~(\ref{Euler}) is the Lorentz force,
which is always normal to the field vector. Note that we consider
non-geodesic worldlines, since the motion of the particles is
dictated by the combined Einstein-Maxwell field and not by gravity
alone. Also, the fluid flow is generally not hypersurface
orthogonal, which explains the presence of the vorticity term in
(\ref{Ray}).

\section{Tolman-Bondi Collapse}\label{break}
The general inhomogeneous collapse of pressure-free spherically
symmetric matter is monitored by means of the Tolman-Bondi metric,
which in the synchronous gauge reads
\begin{equation}
{\rm d}s^2=-{\rm d}t^2+ R'^2(1+k r^2)^{-1}{\rm d}r^2+
R^2{\rm d}\Omega^2\,,  \label{TB}
\end{equation}
where $R=R(r,t)$, $d\Omega^2=d\theta^2+\sin^2\theta d\phi^2$ and a
prime indicates differentiation with respect to $r$
(e.g.~see~\cite{LL}). The spatial curvature index $k=0,\pm1$
corresponds to flat, closed and open geometry respectively. Here
we will only consider the $k=0$  case and throughout this paper we
will use geometrized units with $\kappa=1=c$. In the adopted
coordinate system, the energy momentum tensor of a dust cloud is
$T_{ab}=\rho(r,t)u_au_b$, with $u^a=(1,0,0,0)$ representing the
4-velocity of the fluid. Solved on the above metric Einstein's
equations give
\begin{equation}
R=R(r,t)=\left(\frac{9}{4}\right)^{1/3}M^{1/3}
\left(\tau-t\right)^{2/3} \hspace{15mm} {\rm and} \hspace{15mm}
\rho=\rho(r,t)=\frac{M'}{8\pi R^2R'}\,,  \label{Rrho}
\end{equation}
with the functionals $M=M(r)$ and $\tau=\tau(r)$ describing the
spatial distribution of the matter at fixed time. To proceed
further we use the residual gauge freedom on the radial coordinate
to define $M=\mu r^3/3$, which makes $\mu$ the mass per unit
coordinate volume (i.e. $4\pi r^3/3$). Thus, the only physically
free function left to describe the shape of the pressureless fluid
at a fixed time is $\tau(r)$. Note that the point $t=\tau$
corresponds to a singularity which is reached at a different time
for each shell of fixed radius $r$. Indeed, following
(\ref{Rrho}a) and (\ref{Rrho}b), the Ricci scalar $R^a{}_a=-8\pi
T^a{}_a=8\pi\rho$ associated with any given shell of radius
$r=r^*$, will diverge when $t=\tau(r^*)$. Here, we assume that the
arrow of time increases from $t=0$, which marks the beginning of
the collapse, to $t=\tau(r)$. This in turn guarantees that
$\tau-t>0$ for each $r$. Then, the density and the shear magnitude
of the Tolman-Bondi solution are
\begin{equation}
\rho=\frac{1}{2\pi(\tau-t)\left[3(\tau-t)+2r\tau'\right]}
\hspace{15mm} {\rm and} \hspace{15mm} \sigma^2=\frac{8r^2
\tau'^2}{3[3(\tau-t)+2r\tau']^2(\tau-t)^2}\,,  \label{TBrhosigma2}
\end{equation}
respectively. As we approach the singularity the behavior of these
two variables changes and for $t\rightarrow\tau$ their evolution
is monitored by the following approximate expressions
\begin{equation}
\rho\simeq\frac{1}{4\pi r\tau'(\tau-t)} \hspace{15mm} {\rm and}
\hspace{15mm} \sigma^2\simeq\frac{2}{9\left(\tau-t\right)^2}\,,
\label{sTBrhosigma2}
\end{equation}
indicating that the shear can dominate the final stages of the
collapse. Note that, according to (\ref{sTBrhosigma2}a), we
guarantee a positive definite energy density for the matter by
demanding that $\tau'>0$ at all times. The latter also offers a
sufficient condition for avoiding ¨cross-shell¨ singularities
(see~\cite{LL} for further discussion). Finally, we stress out
that by using the null coordinates, with
\begin{equation}
{\rm d}R=R'{\rm d}r+ \dot R {\rm d}t \hspace{15mm} {\rm and}
\hspace{15mm} {\rm d}v={\rm d}t-(\dot{R}-1)^{-1}{\rm d}R\,,
\label{null}
\end{equation}
the line element (\ref{TB}) reads
\begin{equation}
ds^2=-(1-\dot R^2){\rm d}v^2+ 2{\rm d}v{\rm d}R+ R^2{\rm
d}\Omega^2\,.  \label{TB1}
\end{equation}
Then, the surface where the radial velocities of null congruences
vanish (i.e.~for ${\rm d}R/{\rm d}v=0$) is a spherically symmetric
horizon. Therefore, by putting ${\rm d}s/{\rm d}v={\rm d}R/{\rm
d}v= {\rm d}\Omega/{\rm d}v=0$, the horizon is the surface
$\dot{R}^2=1$. As we are looking for collapsing solutions only
(i.e.~with $\dot{R}<0$), the horizon will be on the restricted
surface $\dot{R}_H=-1$ or, more explicitly, on
$t(r_H)=\tau(r_H)-2\mu r_H^3/9$.

\section{Magnetized Tolman-Bondi Collapse}
Observations have long established the widespread presence of
astrophysical magnetic fields, while compact stellar objects are
capable of supporting considerably strong fields. Neutron stars,
for example, can carry magnetic fields that reach up to
$10^{15}$~G and $10^{16}$~G. Despite their strength, however, the
energy density of these B-fields is much smaller than that of the
supporting matter (i.e.~$B^2/\rho\ll 1$). One can therefore use
its relative weakness to treat the magnetic field as a
perturbation on a matter dominated background. Here, we will
employ the Tolman-Bondi metric to study the collapse of an
inhomogeneous, weakly magnetized dust cloud. We will do so by
adopting the familiar MHD approximation, which is supported by the
expected very high conductivity of stellar interiors.

According to (\ref{Ray}), the volume evolution of a weakly
magnetized fluid element within the irrotational Tolman-Bondi
spacetime is monitored by the following version of the
Raychaudhuri equation
\begin{equation}
\dot{\Theta}+{\textstyle{1\over3}}\Theta^2=
-{\textstyle{1\over2}}\rho- 2\sigma^2+ {\rm D}^a\dot{u}_a+
\dot{u}_a\dot{u}^a\,,  \label{mTBRay}
\end{equation}
given that $p=0$ and $B^2\ll\rho$. The last pair of terms in the
right-hand side of the above is entirely due to the magnetic
presence, since $\dot{u}_a=\epsilon_{abc}B^b{\rm curl}B^c/\rho$
for $p=0$ (see Eq.~(\ref{Euler}) and also~\cite{TB}). In what
follows, we will employ the Tolman-Bondi metric to evaluate these
two terms, while ignoring the magnetic backreaction on the shear
and the vorticity of the background model.

At the MHD limit the electric field vanishes and Maxwell's
equations reduce to a set of one propagation and one constraint
equation. In covariant form, these are given by the respective
expressions
\begin{equation}
h^a{}_b\pounds_u B^b+\Theta B^a=0 \hspace{15mm} {\rm and}
\hspace{15mm}
h^a{}_{b}\nabla_aB^b=0\,,  \label{M}
\end{equation}
where $\pounds_uB^a=u^b\nabla_bB^a-B^b\nabla_bu^a$ is the Lie
derivative of $B^a$ along the 4-velocity of the fluid. Adopting
spherical polar coordinates and recalling that $B_au^a=0$, we set
$B^a=(0,\,B^r,\,B^{\theta},\,B^\phi)$ with
$B^{\alpha}=B^{\alpha}(t,r,\theta,\phi)$ and
$\alpha=r,\,\theta,\,\phi$. Then, solving (\ref{M}a) on our
Tolman-Bondi background we obtain
\begin{equation}
B^{\alpha}=\frac{F_{\alpha}}
{(\tau-t)\left[3(\tau-t)+2r\tau'\right]}\,, \label{sol}
\end{equation}
with $F_{\alpha}$ representing time-independent functionals
(i.e.~$\partial_tF_{\alpha}=0$). Similarly, written relative to a
spherical polar coordinate system, constraint (\ref{M}b)
translates into
\begin{equation}
\sin\theta(rF'_r+2F_r)+r(\sin\theta\partial_\theta F_{\theta}
+\cos\theta F_\theta)=0\,.  \label{div}
\end{equation}

As with the matter density and the shear magnitude before, the
behavior of the magnetic field changes as one gets closer to the
singularity. In particular, for $t\rightarrow\tau$, we find that
the magnitude of the above given $B$-field evolves as
\begin{equation}
B^2\simeq\frac{1}{4}\frac{\tilde{\cal A}F_r^2}{(\tau-t)^{8/3}}\,,
\label{sB2}
\end{equation}
where $B^2=B_{\alpha}B^{\alpha}$ and $\tilde{\cal A}=
{}^3\sqrt{4\mu^2/81}$. This in turn combines with result
(\ref{sTBrhosigma2}a) to provide a measure of the energy-density
ratio $B^2/\rho$ near the Tolman-Bondi singularity
\begin{equation}
\frac{B^2}{\rho}\simeq \frac{{\cal
A}F_r^2r\tau'}{(\tau-t)^{5/3}}\,,  \label{B2/rho}
\end{equation}
with ${\cal A}=\pi\tilde{\cal A}$. Since the magnetic density
grows faster than that of the collapsing matter (compare
(\ref{sB2}) to Eq.~(\ref{sTBrhosigma2}a)), the above also allows
for a rough upper bound on the ratio $B^2/\rho$. It is therefore
clear that the weak-field approximation (i.e.~$B^2/\rho\ll1$) will
hold as long as
\begin{equation}
\tau-t\gg\left({\cal A}F_r^2 r\tau'\right)^{3/5}\,,
\label{cond}
\end{equation}
The $r$-dependence in the right-hand side of this condition
implies that the weak-field approximation can be satisfied at any
time during the contraction. For instance, for any arbitrary
finite value of $\tau-t$, one can always set
$0<r\ll(\tau-t)^{5/3}/{\cal A}F_r^2\tau'$ to ensure that
(\ref{cond}) holds (recall that $\tau'>0$). In other words, we are
always able to identify a shell, labeled by its radius $r$, where
condition (\ref{cond}) is satisfied.

Remaining within the weak-field limit, we will now employ the
Tolman-Bondi metric to evaluate the magnetic impact on the
contracting spacetime. Focusing on the later stages of the
collapse (i.e.~allowing $t\rightarrow\tau$), the dominant
components of last two terms in the right-hand side of the
Raychaudhuri equation (see (\ref{mTBRay})) read
\begin{equation}
D^a\dot{u}_a\simeq -\frac{5}{3}
\frac{{\cal A}F_r^2\tau'}{(\tau-t)^{11/3}} \hspace{15mm} {\rm
and} \hspace{15mm} \dot{u}_a\dot{u}^a\simeq \frac{4}{9}
\frac{{\cal A}F_r^2(\partial_{\theta}F_r)^2\tau^{'2}}
{(\tau-t)^{14/3}}\,,  \label{Dudotu2}
\end{equation}
respectively. Thus, as long as $F_r\neq0$ and
$\partial_{\theta}F_r\neq0$ the former of the above assists the
contraction and the latter acts against it. Note that both terms
grow faster as we approach the Tolman-Bondi singularity, compared
to the matter density and the shear magnitude given by
(\ref{sTBrhosigma2}a) and (\ref{sTBrhosigma2}b) respectively. In
this case the global spherical symmetry of the collapse will be
destroyed by the Lorentz force. Note that, as the supporting
magnetic stress (\ref{Dudotu2}b) is the fastest growing near the
singularity, the overall contribution of the field will tend to
resist the collapse and this could cause the converging worldlines
to bounce. In addition, when cross-shell singularities are not
allowed, halting the gravitational contraction of a shell of
proper radius $r=r_0$ means that all shells with $r>r_0$ will also
cease collapsing. This is possible while still within the
weak-field limit because condition (\ref{cond}) can be satisfied
arbitrarily close to the singularity. In other words, although the
magnetic energy density associated with a given collapsing shell
has negligible contribution to the total energy-momentum tensor,
the Lorentz force dictates the symmetries of the collapse. The
latter results from the generic inhomogeneity of the Tolman-Bondi
spacetime. It is still possible, however, that nonlinear
contributions to the metric, mainly caused by the non-geodesic
motion of the matter, could overwhelm the magnetic resistance and
push the system into further (anisotropic) contraction.

In what follows we will demonstrate that fields with the
aforementioned properties that support against gravitational
collapse can be obtained as solutions of Maxwell's equations on
the Tolman-Bondi background. Indeed, imposing the condition
$\partial_{\theta}F_r\neq0$ on the $B$-field and then solving
Eqs.~(\ref{M}) we arrive at the functionals
\begin{equation}
F_r=-\frac{f}{2}\cos\theta \hspace{15mm} {\rm and} \hspace{15mm}
F_{\theta}=\frac{f}{r}\sin\theta\,,  \label{Fs}
\end{equation}
where $f$ is a constant. Therefore, both components of the
associated magnetic field have a clear $\theta$-dependence. Also,
despite the fact that $B_{\theta}$ diverges at $r=0$, the energy
density of the field is perfectly regular there because
$B^2\propto g_{\theta\theta}F_{\theta}^2$ and
$g_{\theta\theta}\propto r^2$. Following (\ref{Fs}), however, the
radial component of the field and its $\theta$-derivative vanish
at $\theta=\pi/2$ and at $\theta=0,\,\pi$ respectively. Hence,
along these directions the supporting magnetic effect disappears
and the contraction will proceed uninhibited by the presence of
the field.  This behavior, which is probably typical
(see~\footnote{Assuming a solution of (\ref{sol}) with
$\partial_\theta F_r=0$,  we find that $F_{\theta}=[\alpha(r,\phi)
+\beta(r,\phi)\cos\theta]/\sin\theta$. However the latter is
singular at $\theta=0,\,\pi$, which means that $\partial_\theta
F_r$ cannot vanish identically.  Nevertheless, $F_r$ will vanish
for some $\theta\neq0$. In fact, there are always some $\theta$s
such that $\sin\theta\partial_\theta F_\theta+\cos\theta
F_\theta=0$, independently on $r$. For these $\theta$s the only
regular solution (i.e.~non proportional to $1/r^2$ and therefore
non-divergent) of (\ref{div}) is $F_r=0$. Clearly, this proof does
not apply when $\partial_\theta F_\theta=0$. In that case,
however, the solution has $F_r\propto1/\tan\theta$, which diverges
at $\theta=\pi/2$, and it is therefore discarded.}), could be seen
as a direct consequence of the generically anisotropic nature of
the field. The result is an extremely distorted collapse. In
particular, while certain directions will continue collapsing, the
gravitational contraction of most of the magnetized particles will
face strong resistance by the Lorentz force.

At this point we should also emphasize that the
$\theta$-dependence of the radial magnetic component is crucial
for the future of the converging worldlines. To be precise, when
$\partial_{\theta}F_r=0$ the last term in Eq.~(\ref{mTBRay})
approaches the expression
\begin{equation}
\dot u_a\dot u^a\simeq \frac{1}{4} \frac{{\cal
A}F_r^2(F_\theta^2+F_\phi^2\sin^2\theta)} {(\tau-t)^{8/3}}\,,
\label{dotu2}
\end{equation}
instead of (\ref{Dudotu2}a). In this case the later stages of the
magnetized collapse are dominated by (\ref{Dudotu2}a) and
therefore the contraction will proceed unimpeded.

Finally, let us consider the homogeneous limit
$\tau'\rightarrow0$, which corresponds to FRW geometry. In this
special case the shear contribution to the Raychaudhuri vanishes,
while $\rho\propto(\tau-t)^{-2}$ (see Eqs.~(\ref{TBrhosigma2})).
Also, for $\tau'=0$ the right-hand side of both expressions in
(\ref{Dudotu2}) vanish, which means that one needs to evaluate the
magnetic stresses at higher order. Then, the total magnetic effect
near the initial singularity is given by the sum
\begin{equation}
\dot u^a\dot u_a+D^a\dot u_a\simeq K(\tau-t)^{-8/3}\,,
\end{equation}
where $K=K(r,\theta,\phi)$ is an involved function of the
$F_{\alpha}$s and their derivatives and vanishes for an
homogeneous magnetic field. The interesting point is that, as
$t\rightarrow\tau$, the above stress grows faster than the matter
density and therefore it is expected to dominate the final stages
of the collapse. In addition the functional $K$ does not have a
definite sign and therefore its effect on the collapsing dust
depends on the particular magnetic configuration.

\section{Discussion}
Past studies of magnetized gravitational contraction have
indicated that the presence of the field could affect the outcome
of the collapse in nontrivial ways. In~\cite{Th}, for example,
magnetism was found capable of halting the contraction of a finite
cylinder, while in~\cite{AP} the authors provided a static
solution of the Einstein-Maxwell equations with a pressure-free
matter component. It has also been shown that the spherically
symmetric collapse of a charged star could bounce~\cite{novi},
although the bounce seems to cause naked shell-crossing
singularities~\cite{ori}. More recently, it was argued that
magneto-curvature tension stresses could also affect the collapse
of a conventional magnetized fluid~\cite{T}. All these claims
suggest that the Maxwell field could play a key role during the
gravitational collapse of a bounded system. Motivated by that we
have considered the inhomogeneous contraction of a weakly
magnetized Tolman-Bondi spacetime filled with a highly conductive
pressureless fluid. The advantage of the Tolman-Bondi metric is
that it offers the most general mathematical framework for
studying dust collapse. Here, we did so by analyzing the magnetic
contribution to the Raychaudhuri equation, which monitors the
convergence of the particle worldlines in a covariant manner. We
show that the magnetic input splits into two stresses, one of
which always supports against contraction. Assuming that the
energy density of the field is only a small fraction of the matter
density, we have used the Tolman-Bondi metric to evaluate the
input of the aforementioned stresses to Raychaudhuri's formula.
Such an approximation was unavoidable given the fully analytic
nature of our study. Within this limit, we found that the magnetic
presence can severely distort the spherical symmetry of the
collapse. Interestingly, we found that the magnetic stress which
resists the collapse is the fastest growing one and we identified
physically plausible magnetic configurations which allow this to
happen. It should be noted, of course, that by ignoring the
backreaction effects, mainly those due to the non-geodesic motion
of the magnetized fluid, we have limited the range of our results.
Nevertheless, the tendency of the Lorentz force to dominate the
collapse, even when the energy level of the field is relatively
low, should not depend on the approximation level. This argument
is supported by earlier studies, showing that spherically
symmetric charged collapse produces naked
singularities~\cite{ori}. Thus, given that all known stars support
magnetic fields of various strengths, we believe that these fields
can play a protagonist's role in the evolution of such
gravitationally bound systems.\\

\acknowledgments
The authors wish to thank John Barrow and Kostas Kokkotas for helpful
discussions and comments. CG~is supported by a PPARC research grant
(PPA/P/S/2002/00208).

\end{document}